\newcommand{\Wo}{\mbox{${\rm W}_0$}}
\newcommand{\Zsun}{\mbox{${\rm Z}_\odot$}}
\newcommand{\msun}{\mbox{${\rm M}_\odot$}}

\newcommand{\kms}{\mbox{${\rm km~s}^{-1}$}}

\newcommand{\nbody}{\mbox{{{\em N}-body}}}

\newcommand{\tms}{\mbox{${t_{\rm ms}}$}}
\newcommand{\trlx}{\mbox{${t_{\rm rlx}}$}}

\newcommand{\tdf}{\mbox{${t_{\rm f}}$}}
\newcommand{\tdfg}{\mbox{${T_{\rm f}}$}}
\newcommand{\tcc}{\mbox{${t_{\rm cc}}$}}

\newcommand{\tcoll}{\mbox{${t_{\rm coll}}$}}
\newcommand{\tdisr}{\mbox{${t_{\rm disr}}$}}

\newcommand{\tlast}{\mbox{${t_{\rm last}}$}}

\newcommand{\mdf}{\mbox{${m_{\rm f}}$}}

\newcommand{\mm}{\mbox{$\langle m \rangle$}}
\newcommand{\dmcoll}{\mbox{$\langle \delta m \rangle_{\rm coll}$}}
\newcommand{\mmean}{\mbox{${\langle m \rangle}$}}

\newcommand{\Mrunaway}{\mbox{$m_{\rm r}$}}
\newcommand{\mseed}{\mbox{$m_{\rm seed}$}}

\newcommand{\nbf}{\mbox{$n_{\rm bf}$}}
\newcommand{\Ncoll}{\mbox{$N_{\rm coll}$}}

\newcommand{\Rc}{\mbox{${R_{\rm c}}$}}
\newcommand{\Mc}{\mbox{${M_{\rm c}}$}}
\newcommand{\Nc}{\mbox{${N_{\rm c}}$}}
\newcommand{\Vc}{\mbox{${V_{\rm c}}$}}
\newcommand{\Lc}{\mbox{${\Lambda_{\rm c}}$}}

\newcommand{\Rg}{\mbox{${R_{\rm G}}$}}
\newcommand{\Mg}{\mbox{${M_{\rm G}}$}}

\newcommand{\Vg}{\mbox{${V_{\rm G}}$}}
\newcommand{\Lg}{\mbox{${\Lambda_{\rm G}}$}}

\newcommand{\rcore}{\mbox{${r_{\rm core}}$}}

\newcommand{\rhm}{\mbox{${r_{\rm hm}}$}}
\newcommand{\rgc}{\mbox{${R}$}}

\newcommand{\vrot}{\mbox{$v_{\rm rot}$}}
\newcommand{\vorb}{\mbox{$v_{\rm o}$}}

\newcommand{\vdisp}{\mbox{$\langle v^2 \rangle^{1/2}$}}

\newcommand{\ncoll}{\mbox{${\cal N}_{\rm coll}$}}

\newcommand{\fcoll}{\mbox{$f_{\rm coll}$}}
\newcommand{\nrun}{\mbox{${\cal N}_{\rm run}$}}

\newcommand{\tf}{\mbox{${t_{\rm f}}$}}

\newcommand{\erf}{\mbox{${{\rm erf}}$}}

\newcommand{\pc}{\mbox{${\rm pc}$}}

\def\apgt{\ {\raise-.5ex\hbox{$\buildrel>\over\sim$}}\ }
\def\aplt{\ {\raise-.5ex\hbox{$\buildrel<\over\sim$}}\ }

\documentstyle[psfig,aaspp4]{article}

\def\apgt{\ {\raise-.5ex\hbox{$\buildrel>\over\sim$}}\ }
\def\aplt{\ {\raise-.5ex\hbox{$\buildrel<\over\sim$}}\ }
\righthead{Runaway growth in dense star clusters}
\lefthead{S.\ F.\ Portegies Zwart \& S. L. W. McMillan}

\begin{document}


\title{The runaway growth of intermediate-mass black holes in dense
       star clusters}

\medskip 

\author{Simon F.\ Portegies Zwart$^{1, 2}$
	\and
	Stephen L.\ W.\ McMillan$^3$
       }


$^1$ 	Astronomical Institute ``Anton Pannekoek'',
	University of Amsterdam,
	Kruislaan 403,
	1098 SH Amsterdam, NL;
	spz@science.uva.nl

$^2$ 	Section Computational Science, 
	University of Amsterdam,
	Kruislaan 403,
	1098 SH Amsterdam, NL

\bigskip

$^3$	Dept.~of Physics,
	Drexel University, 
	Philadelphia, PA 19104, USA;
	steve@kepler.physics.drexel.edu
\bigskip

\slugcomment{Simon Portegies Zwart is a KNAW Fellow}

Subject headings: 
	stellar dynamics -- 
	binaries (including multiple): close --
	globular clusters: general --
	Galaxies: bulges --
	galaxies: star clusters --
	methods: N-body simulations 


\begin{abstract}
We study the growth rate of stars via stellar collisions in dense star
clusters, calibrating our analytic calculations with direct \nbody\,
simulations of up to 65536 stars, performed on the GRAPE family of
special-purpose computers.
We find that star clusters with initial half-mass relaxation times
$\aplt 25$\,Myr are dominated by stellar collisions,
the first collisions occurring at or near the point of core collapse,
which is driven by the segregation of the most massive stars to the
cluster center, where they end up in hard binaries.
The majority of collisions occur with the same star, resulting in the
runaway growth of a supermassive object.  This object can grow up to
$\sim0.1$\% of the mass of the entire star cluster and could manifest
itself as an intermediate-mass black hole (IMBH).
The phase of runaway growth lasts until mass loss by stellar evolution
arrests core collapse.  Star clusters older than about 5\,Myr and with
present-day half-mass relaxation times $\aplt 100$\,Myr are expected
to contain an IMBH.
\end{abstract}

\section{Introduction}

Using the {\em Chandra} X-ray observatory, Kaaret et al.\, (2000;
2001)\nocite{2001MNRAS.321L..29K}\nocite{2000HEAD...32.1511K} and
Matsumoto et al.\, (2000;
2001)\nocite{2000HEAD...32.0108M}\nocite{2001ApJ...547L..25M} recently
discovered nine bright X-ray sources in the irregular galaxy
M82. Their brightest source (No.\ 7 in Table\,1 of Matsumoto et al.\,
2001) has a luminosity of $9 \times 10^{40} {\rm erg\, s}^{-1}$ in the
0.2--10\,KeV band, corresponding to the Eddington luminosity of a
$\sim 600$\,\msun\, compact object.  The high luminosity and rather
soft X-ray spectrum of the object indicates that it may be an
intermediate-mass black hole (IMBH) with a mass of at least
600\,\msun\, (Kaaret et al.\, 2001; Matsumoto et al.\, 2001).

An optical follow-up in the infrared (J, H, and K$^\prime$-bands) with
the CISCO instrument on the SUBARU telescope revealed a star cluster
with an estimated mass of a few $10^6$\,\msun\, at a position
consistent with the X-ray location of the IMBH (Harashima et~al.\,
2001).  This star cluster appears to be very young ($\aplt 10$\,Myr),
as it is extremely blue and expanding shells of molecular gas have
been discovered in its vicinity (Matsushita et al.\,
2000),\nocite{2000ApJ...545L.107M} typical for a star-forming region
of a few million years.

Matsushita et al.\, (2000) estimate that the environment has an age of
only a few million years.

More unusually bright X-ray point sources have been discovered in the
early spiral galaxies NGC 2403 (Kotoku et al.\,
2000)\nocite{2000PASJ...52.1081K} and NGC 4565 (Mizuno, et
al. 1999)\nocite{1999PASJ...51..663M}. Most remarkable, however, is
the discovery of many bright X-ray sources in the ``Antennae'' system
(NGC 4038/4039) by Fabbiano et
al.\,(2001)\nocite{2001ApJ...554.1035F}, Zazas \& Fabbiano
(2002)\nocite{2002astro.ph..3176Z} and Zazas et al.\,
(2002)\nocite{2002astro.ph..3174Z}, also using {\em Chandra}. These
authors conclude that many of these sources may be $\apgt
100$\,\msun\, accreting black holes (although alternative explanations
exist---see e.g.\, Mizuno 1999; King et al.\,
2001).\nocite{2001ApJ...552L.109K}\nocite{1999AN....320..356M} The
Antennae contain many young star clusters with characteristics similar
to those found in M82 (Mengel et al.\,
2001).\nocite{2001ApJ...550..280M} However, it is not yet clear how
many of the X-ray sources in the Antennae are associated with these
clusters (Zazas \& Fabbiano\, 2002).\nocite{2002astro.ph..3176Z} There
may also be an example of an IMBH in our own Galaxy, as recent
reverberation mapping of the globular cluster M15 by Gebhardt et al.\,
(2000, 2001, and private
communication)\nocite{2000AJ....119.1268G}\nocite{2001AAS...199.5610G}
strongly suggests that the cluster may harbor a $\sim 2500$\,\msun\,
black hole at its center.

Several possible mechanisms for forming IMBHs in star clusters have
recently been suggested.  Miller \& Hamilton (2001) have studied the
possibility that an IMBH may form slowly (on a Hubble time scale) by
occasionally encountering and devouring other cluster stars.  Mouri \&
Taniguchi (2002) have proposed a much more rapid black-hole merger
mechanism, operating in very high-density ($10^6$ black holes
pc$^{-3}$) environments on time scales as short as $\sim10^7$ yr.  In
this paper we consider the possibility of forming a massive object in
a young star cluster due to repeated collisions during an early phase
of core collapse.

Sanders (1970),\nocite{1970ApJ...162..791S} Lee
(1987),\nocite{1987ApJ...319..801L} and Quinlan \& Shapiro
(1990)\nocite{1990ApJ...356..483Q} have studied the possibility of
collision runaways in spherical stellar systems of $\apgt 10^7$ stars
with high ($>100$\kms) velocity dispersions.  All studies began with
stars of equal masses and found that, for sufficiently high densities
and velocity dispersions, runaway mergers could indeed occur.  Quinlan
\& Shapiro observed that the collision time scale for massive
stars decreases faster with increasing mass than does the
main-sequence lifetime, and concluded that clusters with initial
relaxation times of 1--$5 \times 10^8$ years could grow a massive
$\apgt 100$\,{\msun} object by multiple mergers.  Sanders' (1970)
Monte-Carlo calculations neglected the effects of mass segregation and
found collision runaways only after mergers had driven the cluster
into a state of high central density.  However, in the self-consistent
Fokker--Planck models of Lee and Quinlan \& Shapiro, the runaway
started well before core collapse occurred.  All authors concluded
that runaways would not occur in clusters containing less than
$\sim10^6-10^7$ stars because three-body binary heating in small
$N$-systems provided sufficient energy to reverse core collapse before
the runaway process could begin.

In contrast to the studies just described, the models discussed in
this paper begin with a broad range of stellar masses.  Vishniac
(1978)\nocite{1978ApJ...223..986V} demonstrated that a Salpeter
(1955)\nocite{1955ApJ...121..161S} initial mass function is Spitzer
(1969) unstable.\nocite{1969ApJ...158L.139S} As a result, young star
clusters may experience core collapse on the time scale on which the
most massive stars segregate to the cluster center.  This time scale
may be much shorter than the main sequence lifetimes of the stars
involved.  Vishniac suggested that such a prompt collapse might lead
to the formation of a massive compact object.  We find that early core
collapse in a relatively low-$N$ star cluster may result in a
collision runaway, so long as the most massive stars remain on the
main sequence while the collapse occurs.

The possibility of multiple collisions involving the same star in a
dense star cluster was demonstrated convincingly by Portegies Zwart et
al. (1999), using the special-purpose GRAPE-4 (Makino et
al.~1997)\nocite{1997ApJ...480..432M} to speed up their direct N-body
calculations with up to 12288 stars.  They concluded that, even in
small clusters, runaway collisions may lead to the growth of a single
massive star.  The earlier arguments that three-body binary heating
would drive the expansion of the cluster core appear to be unimportant
in these simulations, as mergers between stars tend to destroy
binaries before they can heat the cluster effectively.  Indeed, in
contrast to the underlying assumptions of previous collision studies,
dynamically formed binaries in dense clusters act as a {\em catalyst}
for stellar mergers, boosting the collision rate far beyond the simple
two-body expressions used in earlier work.  The N-body simulations
covered a rather limited part of the available parameter space, but
the initial conditions were selected to mimic known dense star
clusters in the Galaxy and the Large Magellanic Cloud.

If the bright X-ray source in M82 does indeed correspond to a compact
object of $\apgt 600$\,\msun, this IMBH could have been formed by a
collision runaway resulting from collapse of the cluster core early in
the cluster.  In fact, as we will see, it is quite natural to expect a
$\sim 10^3$\,\msun\, black hole in a million-solar-mass star cluster.
We begin by deriving (\S2) some simple analytic expressions describing
the dynamical behavior observed in cluster simulations.  In \S3\, we
calibrate these relations using direct \nbody\, simulations.  We then
(\S4) extend these results to derive simple relations between
black-hole formation and cluster parameters.

\section{Runaway growth of a massive object in a 
	 dense star cluster}\label{sect:ug}

\subsection{Core collapse and the first collision}

A star cluster is a self-gravitating group of stars.  So long as
stellar evolution remains relatively unimportant, the cluster's
dynamical evolution is dominated by two-body relaxation, with
characteristic time scale (Spitzer, 1987)\nocite{1987degc.book.....S}
\begin{equation}
	\trlx = \left( {\Rc^3 \over G \Mc} \right)^{1/2} 
		{\Nc \over 8 \ln\Lc}\,,
\label{Eq:trlx}\end{equation}
the half-mass relaxation time.  Here $G$ is the gravitational
constant, \Mc\, is the total mass of the cluster, $\Nc \equiv
\Mc/\mmean$ is the number of stars and \Rc\, is the characteristic
(half-mass) radius of the cluster.  The Coulomb logarithm $\ln \Lc
\simeq \ln (0.1 \Nc) \sim 10$ typically. In convenient units the
two-body relaxation time becomes
\begin{equation}
	\trlx \simeq 1.9\, {\rm Myr} \left({\Rc \over
		 1\,{\rm pc}}\right)^{3/2} \left({\Mc \over
		1\,\msun}\right)^{1/2} \left( {1\,\msun \over
		\mm}\right) \left(\ln \Lc \right)^{-1}\,.
\label{Eq:trlx_II}\end{equation}
The dynamical evolution of the star cluster drives it toward core
collapse (Antonov 1962; Spitzer \& Hart, 1971)
\nocite{1962spss.book.....A}\nocite{1971ApJ...164..399S}
in which the central density runs away to a formally infinite value in
a finite time. In an isolated cluster in which all stars have the same
mass, core collapse occurs in a time $\tcc
\simeq 15\,\trlx$ (Cohn 1980).
\nocite{1980ApJ...242..765C}\nocite{2001A&A...375..711F} 

Realistic clusters have a broad range in initial stellar masses,
generally from $m_{\rm min} \simeq 0.1$\,\msun\, to $m_{\rm max}
\simeq 100$\,\msun, with mean mass $\mmean$ ranging from
$\sim0.39$\,\msun\, (Salpeter 1955)\nocite{1955ApJ...121..161S} to
about 0.65\,\msun\, (Scalo 1986),\nocite{scalo86} depending on the
specific mass function adopted.  During the early evolution of the
cluster, massive stars sink toward the cluster center via dynamical
friction. Approximating the cluster structure as an isothermal sphere,
we find (Binney \& Tremaine 1987, Eq.\,
7-25)\nocite{1987gady.book.....B} that a star of mass $m$ at distance
$r$ from the cluster center drifts inward at a rate given by
\begin{equation}
	r \frac{dr}{dt} = -0.43 {Gm \over \Vc} \ln \Lc\,,
\label{Eq:tdf_cgs}\end{equation}
Here \Vc\, is the cluster velocity dispersion.  Using
Eq.\,\ref{Eq:trlx_II}, we can integrate Eq.\,\ref{Eq:tdf_cgs} with
respect to time to obtain the dynamical friction inspiral time scale
\begin{equation}
	\tf = 3.3 {\mm \over m} \trlx\,.
\label{Eq:tdf}\end{equation}
This is the time taken for a star of mass $m$ to sink to the cluster
center from a circular orbit at initial distance $r\gg\rcore$.

In a multi-mass system, core collapse is driven by the accumulation of
the most massive stars in the cluster center. This process takes place
on a dynamical friction time scale (Eq.\,\ref{Eq:tdf}).  Empirically,
we find, for initial mass functions of interest here, that core
collapse (actually, the appearance of the first persistent dynamically
formed binary systems) occurs at about
\begin{equation}
	\tcc \simeq 0.20\,\trlx\,.
\end{equation}
This core collapse time is taken in the limit where stellar evolution
is unimportant, i.e.~where stellar mass loss is negligible and the
most massive stars survive until they reach the cluster center.

The collapse of the cluster core may initiate physical collisions
between stars.  The product of the first collision is likely to be
among the most massive stars in the system, and to be in the core.
This star is therefore likely to experience subsequent collisions,
resulting in a collision runaway (see Portegies Zwart et al.\, 1999).
The maximum mass that can be grown in a dense star cluster if all
collisions involve the same star is \Mrunaway, where
\begin{equation}
	{d\Mrunaway \over dt} = \ncoll \dmcoll.
\label{Eq:Mrunaway}\end{equation}
Here \ncoll\, and \dmcoll\, are the average collision rate and the
average mass increase per collision (assumed independent).  We now
discuss these quantities in more detail.  In \S3 we present some
N-body results that both motivated and calibrate the following
discussion.

\subsection{The collision rate \ncoll}

A key result from our simulations is the fact that collisions between
stars generally occur in dynamically formed (``three-body'') binaries.
The collision rate is therefore closely related to the binary
formation rate, which we now estimate.

The flux of energy through the half-mass radius of a cluster during
one half-mass relaxation time is on the order of 10\% of the cluster
potential energy, largely independent of the total number of stars or
the details of the cluster's internal structure (Goodman
1987).\nocite{1987ApJ...313..576G} For a system without primordial
binaries this flux is produced by heating due to dynamically formed
binaries (Makino \& Hut 1990)\nocite{1990ApJ...365..208M}.  It is
released partly in the form of scattering products which remain bound
to the system, and partly in the form of potential energy removed from
the system by escapers recoiling out of the cluster (Hut
\& Inagaki 1985).\nocite{1985ApJ...298..502H} Makino \& Hut argue, for
an equal-mass system, that a binary generates an amount of energy on
the order of $10^2 kT$ via binary--single-star scattering (where the
total kinetic energy of the stellar system is $\frac32\Nc kT$).  This
quantity originates from the minimum binding energy of a binary that
can eject itself following a strong encounter.  Assuming that the
large-scale energy flux in the cluster is ultimately powered by binary
heating in the core. It follows that the required formation rate of
binaries via three-body encounters is
\begin{equation}\label{3bb_rate}
	\nbf \simeq 10^{-3} \frac{\Nc}{\trlx}.
\end{equation}
For systems containing significant numbers of primordial binaries,
which segregate to the cluster core, equivalent energetic arguments
(Goodman \& Hut 1989) lead to a similar scaling for the net rate at
which binary encounters occur in the core.

The above arguments apply to star clusters comprising identical
point-mass stars.  In a cluster with a range of stellar masses,
three-body binaries generally form from stars which are more massive
than average.  After repeated exchange interactions, the binary will
consist of two of the most massive stars in the cluster.  Conservation
of linear momentum during encounters with lower mass stars means that
the binary receives a smaller recoil velocity, making it less likely
to be ejected from the cluster.  The binary must therefore be
considerably harder---$\apgt 10^3 kT$---before it is ejected following
a encounter with another star (see Portegies Zwart \& McMillan
2000).\nocite{2000ApJ...528L..17P}

However, taking the finite sizes of real stars into account, it is
quite likely that such a hard binary experiences a collision rather
than being ejected.  A strong encounter between a single star and a
hard binary generally results in a resonant interaction.  Three stars
remain in resonance until at least one of them escapes, or a collision
reduces the three-body system to a stable binary.  For harder binaries
it becomes increasingly likely that a collision occurs instead of
ejection (McMillan 1986\nocite{1986ApJ...306..552M}).  In the
calculations of Portegies Zwart et al.  (1999) most binaries
experience a collision at a binding energy of order $10^2kT$,
considerably smaller than the binding energy required for ejection.
Accordingly, we retain the above estimate of the binary formation rate
(Eq. \ref{3bb_rate}) and conclude that the collision rate per
half-mass relaxation time is
\begin{equation}
	\ncoll \sim 10^{-3} f_{\rm c} {\Nc \over \trlx}.
\label{Eq:ncoll_binary}\end{equation}
Here we introduce $f_{\rm c} \leq 1$, the effective fraction of
dynamically formed binaries that produce a collision.  Note again that
Eq.\,\ref{Eq:ncoll_binary} is valid only in the limit where stellar
evolution is unimportant.

The most massive star in the cluster is typically a member of the
interacting binary and therefore dominates the collision rate.
Subsequent collisions cause the runaway to grow in mass, making it
progressively less likely to escape from the cluster. The star which
experiences the first collision is therefore likely to participate in
subsequent collisions.  The majority of collisions thus involve one
particular object---the runaway merger---generally selected by its
high initial mass and proximity to the cluster center (see Portegies
Zwart et al.\, 1999).

For systems containing many primordial binaries the above argument
must be modified.  Since dynamically formed binaries tend to be fairly
soft---a few $kT$---we expect that the fraction of interactions with
primordial binaries leading to collision is comparable to the value
$f_{\rm c}$ above.  However, a critical difference is that, in systems
containing many binaries, the collisions involve many different pairs
of stars, not just the binary containing the massive runaway.  For our
proposed runaway scenario to operate, we must assume that high-mass
binaries are rapidly destroyed or merge following interactions in the
core, in which case the above arguments apply.  We note that, once a
runaway begins, binaries have large interaction cross sections and
hence are likely to participate in the runaway process.  From the
point of view of producing massive merger products, the worst-case
scenario would be a substantial primordial population of wide
high-mass binaries.  We are currently carrying out N-body simulations
to investigate the behavior of such systems.

\subsection{Average mass increase per collision}\label{Sect:massincrease}

Once begun, the collision runaway dominates the collision cross
section.  The average mass increase per collision depends on the
characteristics of the mass function in the cluster core.  A lower
limit for stars which participate in collisions can be derived from
the degree of segregation in the cluster.  Inverting Eq.\,\ref{Eq:tdf}
results in an estimate (still assuming an isothermal sphere) of the
minimum mass of a star that can reach the cluster core in time $t$ due
to dynamical friction:
\begin{equation}
	\mdf =	1.9 \msun\,
	        \left({1\,{\rm Myr} \over t}\right)
		\left({\Rc \over 1\,{\rm pc}}\right)^{3/2} 
		\left({\Mc \over 1\,\msun}\right)^{1/2} 
	        \left(\ln \Lc\right)^{-1}\,.
\label{Eq:mdf}\end{equation}
Thus, at time $t$ and for a given mass $m$, there is a maximum radius
$r(t)$ inside of which stars of mass $m$ will have segregated to the
core.  The stars contributing to the growth of the runaway are likely
to be among those more massive than \mdf, because their number density
in the core is enhanced by mass segregation, their collision cross
sections are larger, and they contribute more to $\dmcoll$ when they
do collide.
%


The shape of the central mass function of a segregated cluster is not
trivial to derive.  In thermal equilibrium, the central number
densities of stars of different masses would be expected to scale as
\begin{equation}
	n_0(m) \sim m^{3/2}\,\frac{dN}{dm}\,,
\end{equation}
where $dN/dm$ is the global (Scalo) initial mass function, which
scales roughly as $m^{-2.7}$ at the high-mass end ($m \apgt 10\msun$).
However, as discussed in \S3, the distribution of secondary masses
(i.e.~the masses of the lighter stars participating in collisions)
does not follow the above simple relation.  Rather, we find that stars
in the core do not reach thermal equilibrium (a result generally
consistent with earlier findings by Chernoff and Weinberg
1990\nocite{1990ApJ...351..121C} and Joshi, Nave \& Rasio
2001),\nocite{2001ApJ...550..691J} and that the dynamical nature of
the collisional processes involved mean that more massive stars tend
to be consumed before lower-mass stars arrive in the core.  In
addition, most collisions involve three-body binary formation and
binary interactions in a multi-mass environment, further complicating
the connection between stellar densities and secondary masses.

Empirically, we find that the secondary mass distribution is quite
well fit by a power-law, $dN/dm \propto m^{-2.3}$ (coincidentally very
close to a Salpeter distribution).  Integrating this expression from a
minimum mass of {\mdf} (and ignoring the upper limit) results in a
mean mass increase per collision of
\begin{equation}
	\dmcoll \simeq 4 m_{\rm f}\,.
\label{Eq:mcoll}\end{equation}
We neglect stars with masses less than $\mdf$.  Substitution of
Eq.\,\ref{Eq:trlx} into Eq.\,\ref{Eq:mdf} and Eq.\,\ref{Eq:mcoll} then
results in a mass increase per collision of
\begin{equation}
	\dmcoll \simeq 4 {\trlx \over t} \mm \ln \Lc\,.
\label{Eq:dmcoll}\end{equation}
Taken over the entire ``collisional'' lifetime of the core, it is
perhaps not surprising that the net distribution of secondary masses
tends to follow the overall distribution of high-mass stars.

\subsection{Lifetime of a cluster in a static tidal field}

With simple expressions for $\ncoll$ and $\dmcoll$ now in hand, we
return to the determination of the runaway growth rate
(Eq.~\ref{Eq:Mrunaway}).  The evaporation of a star cluster which
fills its Jacobi surface in an external potential is driven by tidal
stripping.  Portegies Zwart et al (2001a) have studied the evolution
of young compact star clusters within $\sim200$\,pc of the Galactic
center.  Their calculations employed direct {\nbody} integration,
including the effects of both stellar and binary evolution and the
(static) external influence of the Galaxy, and made extensive use of
the GRAPE-4 (Makino et al.\, 1997) special-purpose computer.  They
found that the mass of a typical model cluster decreased almost
linearly with time:
\begin{equation}
	\Mc = M_{\rm c0} \left(1 - \frac{t}{\tdisr}\right)\,.
\label{Eq:mass}\end{equation}
Here $M_{\rm c0}$ is the mass of the cluster at birth and \tdisr\, is
the cluster's disruption time.  Portegies Zwart et
al.~(2001a)\nocite{2001ApJ...546L.101P} found that their model
clusters dissolved within about 30\% of the two-body relaxation time
at the tidal radius (defined by substituting the tidal radius instead
of the virial radius in Eq.\,\ref{Eq:trlx}).  In terms of the
half-mass relaxation time, we find $\tdisr = 1.6$--5.4\,\trlx,
depending on the initial density profile (the range corresponds to
King [1966]\nocite{1966AJ.....71...64K} dimensionless depths
$\Wo=3$--7; more centrally condensed clusters live longer).

Substituting Eqs.\,\ref{Eq:ncoll_binary} and \ref{Eq:dmcoll} into
Eq.\,\ref{Eq:Mrunaway}, and defining $M_{\rm c0} = \Nc\mmean$ to rewrite
Eq.\,\ref{Eq:mass} in terms of the number of stars in the cluster,
we find
\begin{eqnarray}
	{d\Mrunaway \over dt} 
		&=& 4 \times 10^{-3} f_{\rm c}
				{\Nc \mmean \ln \Lc \over t} 
			\nonumber \\
		&=& 4 \times 10^{-3} f_{\rm c} M_{\rm c0} \ln \Lc
				\left(\frac{1}{t} - \frac{1}{\tdisr}\right)\,.
\end{eqnarray}
Integrating from $t=\tcc$ to $t=\tdisr$ results in
\begin{equation}
	\Mrunaway = \mseed + 4 \times 10^{-3} f_{\rm c}
		     M_{\rm c0} \ln \Lc
			\left[ 
      			     \ln \left( {\tdisr \over \tcc} \right) 
				+ {\tcc \over \tdisr} - 1
			\right] \,.
\label{Eq:mrunaway}\end{equation}
Here $\mseed$ is the seed mass of the star which initiates the runaway
growth, most likely one of the most massive stars initially in the
cluster.  With $\tcc \simeq 0.2\trlx$, Eq\,\ref{Eq:mrunaway} reduces to
\begin{equation}
	\Mrunaway = \mseed + 4 \times 10^{-3} f_{\rm c}
			 M_{\rm c0} \gamma \ln \Lc \,,
\label{Eq:Mrunaway_II}\end{equation}
where $\gamma \simeq \ln \tdisr/\tcc + \tcc/\tdisr - 1 \sim 1$.

\section{Results of {\nbody} simulations}\label{sect:simulations}

The development of the GRAPE family of special-purpose computers makes
it relatively straightforward to test and tune the above simple model
using direct {\nbody} calculations.  Tab.\,\ref{Tab:models} summarizes
the results of an extensive series of detailed {\nbody} simulations of
core collapse and stellar collisions in dense star clusters containing
up to 65536 stars.  These simulations were performed using the
``Starlab'' software environment (see Portegies Zwart et
al.~2001b)\nocite{2001MNRAS.321..199P} running on the GRAPE-6 (Makino
2000).\nocite{2001dscm.conf...87M} The calculations were performed
with initially single stars but, as just discussed, the presence of
primordial binaries is not likely to change the picture qualitatively.
To expand on these findings, we have performed an additional series of
simulations with $\sim10^4$ stars using the same software and
hardware.  Further simulations of systems containing substantial
numbers of primordial binaries are in progress, but are much more time
consuming, due to the complexity of following binary and multiple
encounters in a large-$N$ context.

\subsection{Core collapse}

In our isolated star clusters (three calculations) with $10^4$
identical single point masses distributed as a Plummer model, core
collapse occurs at $\tcc \simeq 15.2\pm 0.1\,\trlx$.  This result is
consistent with earlier calculations of e.g., Cohn (1980) and Makino
(1996).\nocite{1996ApJ...471..796M} Doubling the mass of 20\% of the
stars reduced the core collapse time to $\tcc \simeq 7.2\,\trlx$.
Making 20\% of the stars 10 or 100 times more massive reduced the time
of core collapse further, to $\tcc \simeq 1.4\,\trlx$ and $\tcc\simeq
0.16\,\trlx$, respectively.

The more realistic models of Portegies Zwart et al.\,
(1999)\nocite{pzmmh99} with 6144 and 12288 single stars drawn from a
Scalo (1986) initial mass function also include mass loss from stellar
evolution.  The initial density distributions for these models were
$\Wo=6$ King (1966) models.  Core collapse in these models occurred at
$\tcc \simeq 0.19\pm0.08\,\trlx$.  The slightly later collapse
compared to the models just described, containing $10^4$ identical
point masses and a heavy component, may be attributable to the rather
different mass function, as well as to stellar mass loss, which tends
to delay core collapse.

\subsection{Collision rate}

Relaxing the assumption of point masses to include finite stellar
sizes introduces collisions into our models.  In all calculations, the
first collision occurred shortly after the formation of the first
$\apgt 10\,kT$ binary by a three-body encounter, i.e. close to the
time of core collapse.  When stars were given unrealistically large
radii (100 times larger than normal), the first collisions occurred
only slightly (about 5\%) earlier.

\begin{figure}[htbp!]
\hspace*{1.cm}

\psfig{figure=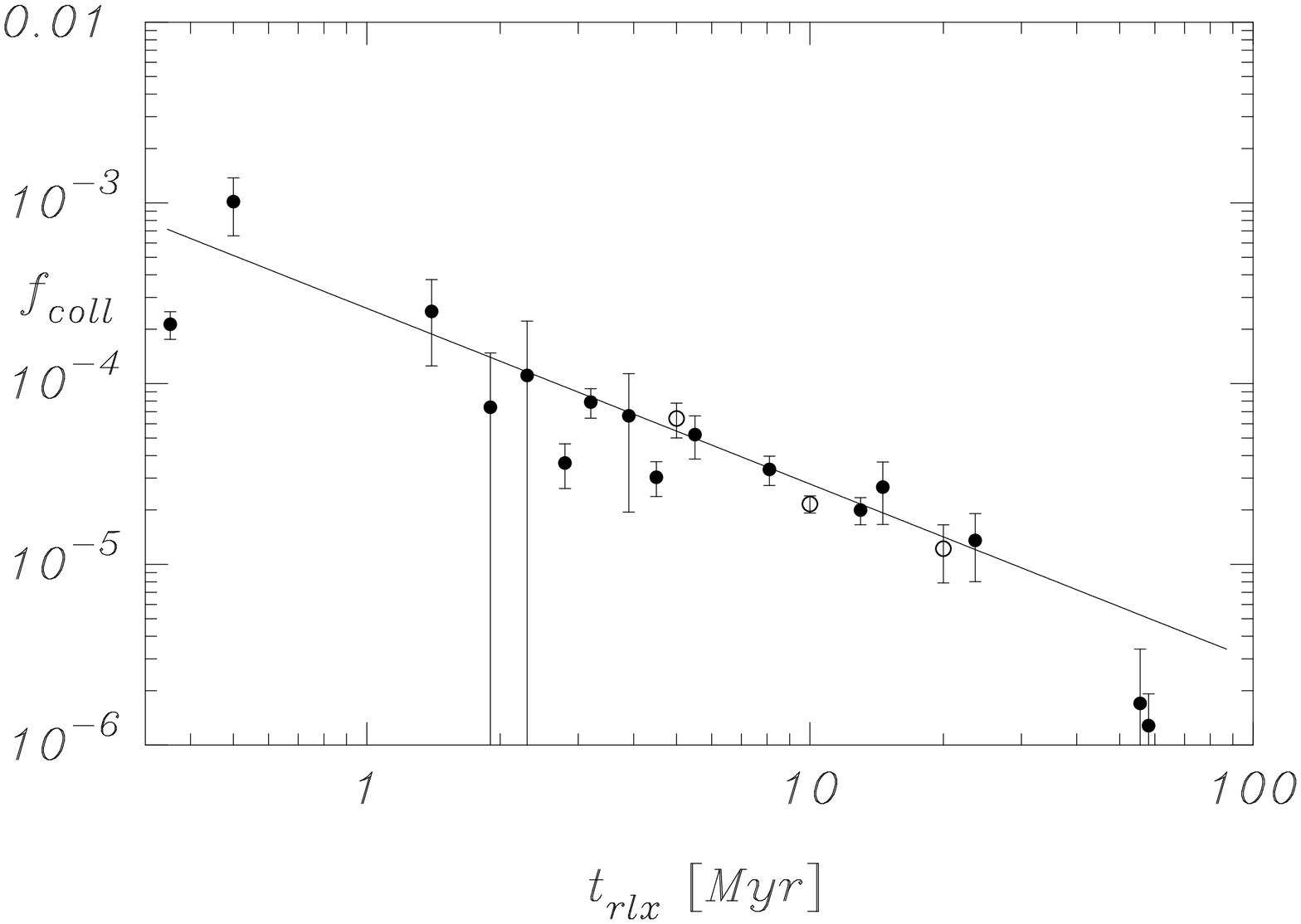,width=10.cm,angle=-90}

\caption[]{Mean collision rate $\fcoll = {\Ncoll / \Nc \tlast}$ as
function of initial relaxation time for all models of
Tab.\,\ref{Tab:models}.  Here \tlast\, is the time of the last
collision in the cluster. The open circles give the results of systems
which are isolated from the Galactic potential (see Portegies Zwart et
al 1999).  Vertical bars represent Poissonian 1-$\sigma$ errors.  The
solid line is a least squares fit to the data (see
Eq.\,\ref{Eq:ncoll}).  The strong reduction in the collision rate for
cluster with an initial relaxation time $\trlx \apgt 30$\,Myr is
probably real.}
\label{Fig:ncoll_Trlx}
\end{figure}

As discussed earlier, the first star to experience a collision was
generally one of the most massive stars in the cluster; this star then
became the target for further collisions.  In models with initial
relaxation times greater than about 30\,Myr the target star exploded
in a supernova before experiencing runaway growth.  The collision
rates in these clusters were considerably smaller than for clusters
with smaller relaxation times (see Fig.\,\ref{Fig:ncoll_Trlx}).
As discussed in more detail in \S4, the onset of stellar evolution
terminates the collision process; premature disruption of the cluster
also ends the period of runaway growth.

\begin{table}[htbp!]
\caption[]{
Overview of the {\nbody} calculations on which the collision rates
reported in this paper are based.  The first column gives the name of
the model, as defined in previous publications (the names RxWx and
KMLx are from Portegies Zwart et al. 2001b [see also Portegies Zwart
et al. 2001a]; the other models are described in detail by Portegies
Zwart et al 1999).\nocite{pzmmh99} The next five columns give the
number of stars (in units of 1024), the initial mass function (Scalo
1986, or a power law with slope as indicated), the initial King
parameter \Wo, the initial relaxation time (in Myr), and the number of
runs performed with these initial conditions (\nrun).  The final three
columns give the average number of collisions in these calculations,
the moment the last collision occurred, and the mean collision rate
per Myr per star.  The models indicated with $\star$ were computed
without a Galactic tidal field (see Portegies Zwart et al.~1999).}
\begin{flushleft}
\begin{tabular}{llclrlrrr} \hline
model&
$\langle N\rangle$&
IMF&
$\langle \Wo\rangle$&
$\langle \trlx\rangle$&
\nrun&
$\langle \Ncoll \rangle$&
$\langle \tlast \rangle$&
\fcoll \\
\hline
R34W7  &   12k &  Scalo&   7& 0.4&    2&  16.&    10.4& -3.90  \\
KML112 &    4k &  -2   &   7& 0.5&    2&  4.0&    1.9 & -3.29  \\
KML101 &    4k &  -2   &   4& 1.4&    2&  2.0&    1.0 & -3.31 \\  
KML142 &    6k &  -2.35&   4& 1.9&    1&  1.0&    2.2 & -4.13  \\
KML111 &    4k &  -2   &   1& 2.3&    2&  0.5&    6.7 & -4.74 \\
R90W7  &   12k &  Scalo&   7& 2.8&    1&  13.&    10.0& -3.98  \\
N64R6r36&  64k &  Scalo&   3& 3.2&    1&  10.&     1.0& -3.82  \\
R34W4  &   12k &  Scalo&   4& 3.2&    3&  6.3&    30.0& -4.71  \\
KML144 &   14k &  -2.35&   4& 3.9&    1&  2.0&    2.4 & -4.24  \\
R150W7 &   12k &  Scalo&   7& 4.5&    2&  10.&    21.3& -4.42  \\
6k6X5$^\star$  		    
       &    6k &  Scalo&   6& 5.0&    1&  21.&    47.9& -4.15  \\
R34W1  &   12k &  Scalo&   1& 5.5&    3&  4.7&    29.1& -4.88  \\
R90W4  &   12k &  Scalo&   4& 8.1&    5&  5.8&    10.0& -4.33  \\
Nk6X10$^\star$ 		    
       &    9k &  Scalo&   6&10.0&    8&  10.&    18.0& -4.22  \\
R150W4 &   12k &  Scalo&   4&13.0&    4&  8.5&    7.3 & -4.02  \\
R90W1  &   12k &  Scalo&   1&14.6&    1&  7.0&    9.8 & -4.24  \\
6k6X20$^\star$ 		    
       &    6k &  Scalo&   6&20.0&    2&  4.0&    95.4& -5.17  \\
R150W1 &   12k &  Scalo&   1&23.6&    2&  3.0&    2.1 & -3.93  \\
R300W4 &   12k &  Scalo&   4&55.6&    1&  1.0&    10.0& -5.09  \\
R34W1  &   32k &  Scalo&   1&58.1&    1&  4.0&    35.8& -5.47  \\
\hline     		       
\end{tabular}
\end{flushleft}
\label{Tab:models} \end{table}

The 45 $N$-body calculations listed in Tab.\,\ref{Tab:models} span a
broad range of initial conditions.  The number of stars varied from 1k
(1024) to 64k (65536).  Initial density profiles and velocity
dispersion for the models were taken from Heggie-Ramamani models
(Heggie \& Ramamani 1995)\nocite{1995MNRAS.272..317H} with {\Wo}
ranging from 1 to 7.  At birth, the clusters were assumed to fill
their zero-velocity (Jacobi) surfaces in the Galactic tidal field.  In
most cases we adopted an initial mass function between 0.1\,\msun\ and
100\,\msun\, suggested for the Solar neighborhood by Scalo
(1986).\nocite{scalo86} However, several calculations were performed
using power-law initial mass functions with exponents of -2 or -2.35
(Salpeter) and lower mass limits of 1\,\msun.  The model with 64k
stars (model N64R6r36) was initialized with a Scalo (1986) mass
function, but with a lower mass limit of 0.3\,\msun\, instead of the
0.1\,\msun\, used in the other models.  The characterization of the
tidal field is discussed in Portegies Zwart et al.\,(2001b).

The number of collisions in these simulations ranged from 0 to 24.
Fig.\,\ref{Fig:ncoll_Trlx} shows the mean collision rate {\ncoll} per
star per million years as a function of the initial half-mass
relaxation time.  The solid line in Fig.\,\ref{Fig:ncoll_Trlx} is a
fit to the simulation data, and has
\begin{equation}
 	\ncoll = 2.2\times 10^{-4} {\Nc \over \trlx}\,,
\label{Eq:ncoll}\end{equation}
for $\trlx\aplt 20-30$\,Myr, consistent with our earlier estimate
(Eq.\,\ref{Eq:ncoll_binary}) if $f_{\rm c} = 0.2$.  The quality of the
fit in Fig.\,\ref{Fig:ncoll_Trlx} is quite striking, especially when
one bears in mind the rather large spread in initial conditions for
the various models.

\begin{figure}[htbp!]
\hspace*{1.cm}

\psfig{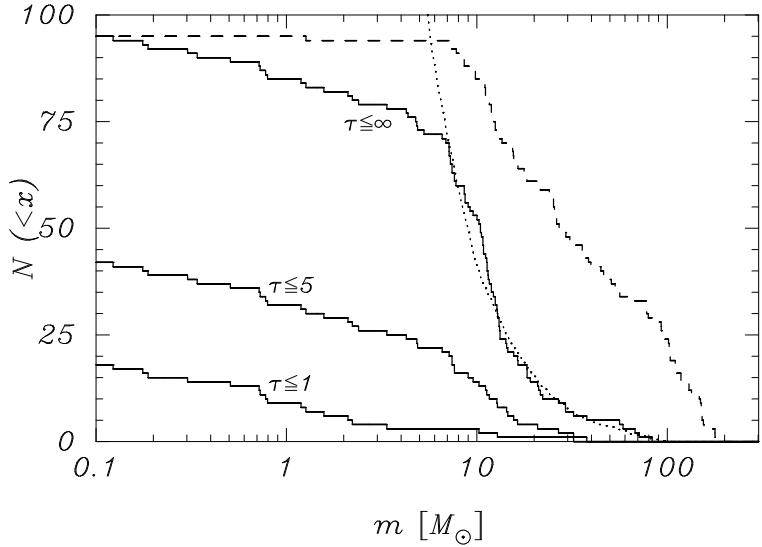}

\caption[]{Cumulative mass distributions of primary (dashed line) 
and secondary (solid lines) stars involved in collisions.  Only those
secondaries experiencing their first collision are included.  From
left to right, the solid lines represent secondary stars for which $\tau
\equiv \tcoll/\tdf \aplt 1$, 5, and $\infty$.  The numbers of
collisions included in each curve are 18 (for $\tcoll/\tdf \aplt 1$),
42 and 95 (rightmost two curves).  The dotted line gives a Power-law
fit with the Salpeter exponent (between 5\,\msun\, and 100\,\msun) to
the right most solid curve ($\tau \aplt \infty$).  }
\label{Fig:dmcoll_distrn}
\end{figure}

Figure\,\ref{Fig:dmcoll_distrn} shows the cumulative mass
distributions of the primary (more massive) and secondary (less
massive) stars participating in collisions.  We include only events in
which the secondary experienced its first collision (that is, we omit
secondaries which were themselves collision products).  In addition,
we distinguish between collisions early in the evolution of the
cluster and those that happened later by subdividing our data based on
the ratio $\tau = \tcoll/\tf$, where $\tcoll$ is the time at which a
collision occurred and $\tf$ is the dynamical friction time scale of
the secondary star (see Eq.\,\ref{Eq:tdf}).  The solid lines in
Figure\,\ref{Fig:dmcoll_distrn} show cuts in the secondary masses at
$\tau \aplt 1$, $\tau \aplt 5$ and $\tau < \infty$ (rightmost line).
The mean secondary masses are $\mmean = 4.0\pm 4.8$\,\msun,
$8.2\pm6.5$ and $\mmean = 13.5\pm8.8$\,\msun\, for $\tau
\aplt 1$, 5 and $\infty$, respectively.

The distribution of primary masses in Figure\,\ref{Fig:dmcoll_distrn}
(dashed line) hardly changes as we vary the selection on $\tau$.  We
therefore show only the full ($\tau \aplt \infty$) data set for the
primaries.  In conrast, the distribution of secondary masses changes
considerably with increasing $\tau$.  For small $\tau$, secondaries
are drawn primarily from low-mass stars.  As $\tau$ increases, the
secondary distribution shifts to higher masses while the low-mass part
of the distribution remains largely unchanged.  The shift from
low-mass ($\aplt 8$\,\msun) to high-mass collision secondaries ($\apgt
8$\,\msun) occurs between $\tau=1$ and $\tau=5$.  This is consistent
with the theoretical arguments presented in
Sec.\,\ref{Sect:massincrease}.  During the early evolution of the
cluster ($\tau\aplt 1$), collision partners are selected more or less
randomly from the available (initial) population in the cluster core;
at later times, most secondaries are drawn from the mass-segregated
population.

Interestingly, although hard to see in Fig.\,\ref{Fig:dmcoll_distrn},
all the curves are well fit by power laws between $\sim 8$\,\msun\,
and $\sim 80$\,\msun (0.8\,\msun\, and 30\,\msun for the leftmost
curve).  The power-law exponents are $-0.4$, $-0.5$ and $-2.3$ for
$\tau \aplt 1$, 5, and $\infty$, and $-0.3$ for the primary (dashed)
curve.  (Note that the Salpeter mass function has exponent $-2.35$.)

Figure\,\ref{fig:bhmass} shows the maximum mass of the runaway
collision product as function of the initial mass of the star cluster.
Only the left side ($\log M/\msun \aplt 7$) of the figure is relevant
here; we discuss the extrapolation to larger masses in
Sec.\,\ref{Sect:massive_black_holes}.  The N-body results are
consistent with the theoretical model presented in
Eq.\,\ref{Eq:mrunaway}.

\section{Discussion}\label{sect:discussion}

Early core collapse in dense star clusters may initiate a phase of
runaway stellar growth, leading to an object containing up to $\sim
0.1$\,\% of the total cluster mass.  We do not address here the state
of this object, which could be a black hole or a star.  If the object
is a helium- or hydrogen-burning star, it may collapse into a compact
object when it exhausts its central fuel.  The amount of mass lost in
the supernova explosion and whether the compact object receives a
velocity kick are important considerations for the future evolution of
the collision runaway.  An extensive parameter study of the details of
the supernova is beyond the scope of this paper.  The basis of our
analysis, however, is simple and robust to quite substantial
perturbations.

We now consider the circumstances under which runaway growth may be
prevented or terminated at an early stage.  Premature termination of
the runaway occurs when stellar mass loss starts to drive the
expansion of the star cluster, or when the star cluster is disrupted
by external influences.  At the end of this section
(Sec.\,\ref{Sect:massive_black_holes}) we briefly discuss the
application of runaway growth to the possible formation of
supermassive black holes in the bulges of galaxies.

\subsection{Prevention of the collision runaway}

Runaway growth in a star cluster can only occur when stellar evolution
is relatively unimportant compared to the dynamical evolution of the
cluster.  Stellar mass loss tends to heat the cluster by loss of
potential energy, and can easily reverse core collapse.  This is
particularly true for the most massive stars, which dominate the
dynamics of the cluster core and are also the first to lose
substantial amounts of mass in stellar winds and supernovae.  The
prevention of core collapse also prevents the first or any subsequent
collisions, and a reversal of core collapse terminates the collision
runaway.  As a rule of thumb, we argue that runaway growth can be
prevented when the time scale for the most massive stars to segregate
to the cluster center exceeds the lifetimes of those stars.

The main-sequence lifetime (\tms) for stars more massive than
$\sim25$\,\msun\, is a rather flat function of mass.  For high
metallicities, the most massive stars ($m\apgt 85$\,\msun) actually
live longer than stars with masses between 25 and $65 \msun$ (Meynet
et al.\, 1994).\nocite{1994A&AS..103...97M} For $Z=0.04$ (recall that
the solar metallicity is $\Zsun = 0.02$), the hydrogen plus helium
burning lifetime varies from 6.27\,Myr for a $25\msun$ star to
7.57\,Myr for a $120\msun$ star.  For low metallicities ($Z=0.001$),
the range becomes 8.19--3.20\,Myr for the same masses (see Meynet et
al. for details).

For the star cluster to experience core collapse before the most
massive stars evolve, we require $\tcc \simeq 0.2\trlx \aplt
\tms(120\,\msun) \sim 3.20$---$7.57$\,Myr for $Z=0.001$--0.04. Runaway
growth therefore does not occur in star clusters with initial
relaxation times $\trlx \apgt 16$--38\,Myr.  For definiteness, we
conclude that clusters with $\trlx \aplt 25$\,Myr experience core
collapse before the most massive stars explode, and are therefore
prone to runaway collision.

The half-mass radius of a tidally limited cluster expands during core
collapse, causing the mean relaxation time to increase by about a
factor of 4 (see Portegies Zwart et al.\, 2001).  A cluster with an
initial relaxation time of $\trlx \simeq 25$\,Myr will therefore have
a relaxation time of about 100\,Myr after core collapse.  Such a
cluster will not experience any further collision runaway, but may
still contain the evidence of such a phase in the form of a central
compact object with a mass $\aplt 0.1$\% of the initial cluster mass.
The cluster may also be relatively depleted in low-mass compact
objects (stellar mass black holes and neutron stars), as these are
consumed during the runaway growth phase (Portegies Zwart et al.\,
1999).

\subsection{Early termination of the runaway by tidal
disruption}\label{Sect:terminate}

A star cluster in orbit around the Galactic center is subject to
dynamical friction, in much the same way as dynamical friction drives
massive stars toward the cluster center. This causes the cluster to
spiral into the Galactic center, where it is destroyed (see Gerhard
2001).\nocite{2001ApJ...546L..39G} We derive here in some detail the
dynamical friction time scale for a star cluster in the potential of
the Galactic center.  We assume constant cluster mass $M_c$, deferring
the more realistic case of a time-dependent cluster mass (cf.\,
Eq.\ref{Eq:mass}) to McMillan \& Portegies Zwart (2002, in
preparation).

The drag acceleration due to dynamical friction is (equation [7-18] in
Binney \& Tremaine, 1987)\nocite{1987gady.book.....B}
\begin{equation}
	a = -{4\pi \ln\Lg G^2 \Mc \rho_G(\Rg) \over \vorb^2}
	\left[
	\erf(X) - {2X \over \sqrt{\pi}} e^{-X^2} \right]\,.
\label{Eq:Fdf}\end{equation}
Here $\ln \Lg$ is the Coulomb logarithm for the Galactic central
region, for which we adopt $\ln \Lg \sim \Rg/\Rc$, $\erf$ is the error
function and $X \equiv
\vorb/\sqrt{2}\Vg$, where $\Vg$ is the one-dimensional velocity
dispersion of the stars at distance \Rg\, from the Galactic center.

The mass of the Galaxy lying within the cluster's orbit at distance
{\Rg} ($\aplt500$\,pc) from the Galactic center is (Sanders \&
Lowinger 1972; Mezger et
al.\,1996)\nocite{1996A&ARv...7..289M}\nocite{1972AJ.....77..292S}
\begin{equation}
	\Mg(\Rg) = 4.25 \times 10^6 \left({\Rg \over 1\,{\rm pc}}
		     	           \right)^{1.2}           \;\;\msun\,.
\label{Eq:Mgal}\end{equation}
Its derivative, the local Galactic density (see Portegies Zwart et
al.\, 2001) is
\begin{equation}
	\rho_G(\rgc) \simeq 4.06\  \times 10^5 \left({\Rg \over 1\,{\rm pc}}
		     	           \right)^{-1.8}  \;\;\msun\,{\rm pc}^{-3}\,.
\label{Eq:rhogal}
\end{equation}
For inspiral through a sequence of nearly circular orbits, the
function $\erf(X) - {2X \over \sqrt{\pi}} \exp(-X^2)$ appearing in
Eq. \ref{Eq:Fdf} may be determined as follows.

Following Binney \& Tremaine (p.\,226), we write the equation of
dynamical equilibrium for stars near the Galactic center as
\begin{equation}
	\frac{dP}{d\Rg} = - \rho_G \frac{G\Mg(\Rg)}{\Rg^2}\,,
\label{Eq:hydro_eq}
\end{equation}
where $P = kT\rho/\mmean$, $\frac32kT = \frac12 \mmean\langle
v^2\rangle$. Since $\sigma^2 =
\frac13\langle v^2\rangle$, it follows that $P =
\sigma^2\rho$, and Eq. \ref{Eq:hydro_eq} becomes
\begin{equation}
	\frac{d~}{dr}(\sigma^2\rho) = -\frac{\rho}r\,\vorb^2\,,
\end{equation}
where \vorb\, is the circular orbital velocity at radius $R$: $\Vg^2 =
G\Mg(\Rg)/\Rg$.  For $\Mg \propto \Rg^x$ (see Eq.\,\ref{Eq:Mgal}), and
assuming that $\Vg^2\propto \vorb^2\sim
\Rg^{x-1}$, we find $\Vg^2\rho \sim \Rg^{2x-4}$, so
\begin{equation}
	r\,\frac{d~}{dr}(\Vg^2\rho) = (2x-4)\Vg^2\rho
			= -\rho \vorb^2\,,
\end{equation}
and hence $X = \sqrt{2-x}$.  Eq. \ref{Eq:Fdf} then becomes
\begin{equation}
	a = -1.2 \ln\Lg \frac{G \Mc }{\Rg^2} \; 
		\left[ \erf(X) - {2X \over \sqrt{\pi}} \exp(-X^2) \right]\,.
\label{Eq:Fdf_II}\end{equation}
For $x = 1.2$, $X= 0.89$ and
\begin{equation}
	a = -0.41 \ln\Lg \frac{G\Mc}{\Rg^2}\,.
\label{Eq:Fdf_IIa}\end{equation}

Again following Binney \& Tremaine, defining $L = \Rg \vorb$ and
setting $dL/dt = a\Rg$, we can integrate Eq.\,\ref{Eq:Fdf_IIa} with
respect to time to find an inspiral time from initial radius $R_i$
of
\begin{eqnarray}
	\tdfg &\simeq & \frac{1.28}{\ln\Lg}
			\frac{\Mg(R_i)}{\Mc}
			\left[\frac{G\Mg(R_i)}{R_i^3}\right]^{-1/2} \\
	&\simeq& 1.4
		\left(\frac{R_i}{10 {\rm pc}}\right)^{2.1}
		\left(\frac{10^6 \msun}{\Mc}\right) \; {\rm Myr}
		\label{Eq:tdf_GC}\label{Eq:tdfapprox}
\end{eqnarray}
For definiteness, we have assumed $\ln\Lg\sim 4$ ($\Lg \sim \Rg/\Rc
\sim 100$) in Eq. \ref{Eq:tdf_GC}, corresponding to a distance of
about 10--30\,pc from the Galactic center.



The maximum mass of the runaway merger for clusters which are
disrupted by inspiral (which of course always destroys the cluster
before it reaches the center) may be calculated by replacing $\tdisr$
in Eq.\,\ref{Eq:mrunaway} by $\tdfg$.  The right-hand side of that
equation then becomes a function of
\begin{equation}
	\frac{\tdfg}{\tcc}
		\simeq 9.0
		\left( {R_i \over  10 \pc} \right)^{2.1} 
		\left( {0.25 \pc \over \Rc} \right)^{3/2}
		\left( {10^5 \msun \over \Mc} \right)^{3/2}
\end{equation}
We can also estimate the maximum initial distance from the Galactic
center for which core collapse occurs (and hence runaway merging may
begin) before the cluster disrupts by setting $\tdfg = \tcc$.  The
result is $R_i \apgt 0.0025 \pc \left(\Rc\Mc/[\pc\,\msun]
\right)^{0.71}$.  For $R_c = 0.25$\,pc and $M_c = 10^5 \msun$, we find
$R_i \apgt 3.3$\,pc. 

\subsection{Speculation on the formation of supermassive black holes}
\label{Sect:massive_black_holes}

A million-solar-mass star cluster formed at a distance of $\aplt
30$\,pc from the Galactic center can spiral into the Galactic center
by dynamical friction before being disrupted by the tidal field of the
Galaxy (see Gerhard 2001).\nocite{2001ApJ...546L..39G} Only the
densest star clusters survive to reach the center.  These clusters are
prone to runaway growth and produce massive compact objects at their
centers.  Upon arrival at the Galactic center, the star cluster
dissolves, depositing its central black hole there.  Black holes from
in-spiraling star clusters may subsequently merge to form a
supermassive black hole.  Ebisuzaki et al.~(2001) have proposed that
such a scenario might explain the presence of the central black hole
in the Milky Way galaxy.

If we simply assume that bulges and central supermassive black holes
are formed from disrupted star clusters, this model predicts a
relation between black hole and bulge masses in galaxies similar to
the expression (Eq.\,\ref{Eq:Mrunaway_II}) connecting the mass of an
IMBH to that of its parent cluster.  However, the ratio of stellar
mass to black-hole mass might be expected to be smaller for galactic
bulges than for star clusters, because not all star clusters produce a
black hole and not all star clusters survive until the maximum black
hole mass is reached.  We would expect, however, that the general
relation between the black hole mass and that of the bulge remains
valid.

\begin{figure}[ht]

\psfig{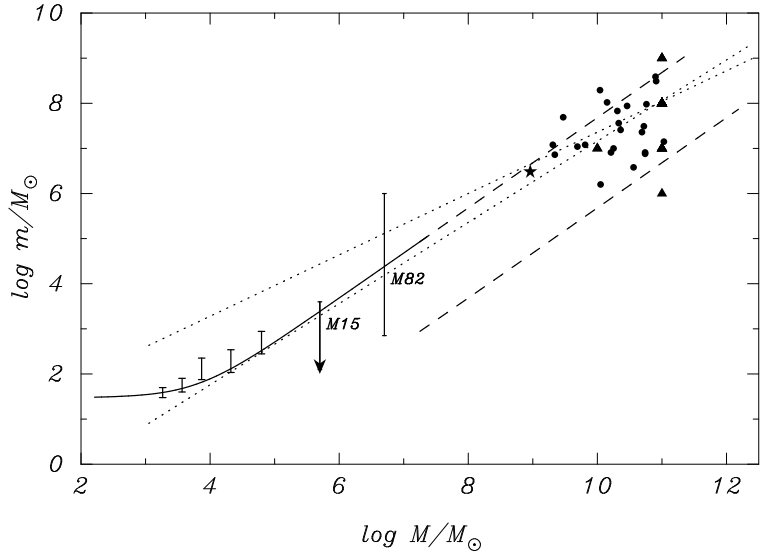}

\caption[]{The mass after a period of runaway growth as a function of
the mass of the star cluster.
The solid line is $\Mrunaway = 30 + 8 \times 10^{-4} M_{\rm c0} \ln
\Lc$ (see Eq.\,\ref{Eq:Mrunaway_II} with $f_{\rm c}=0.2$, $\gamma=1$
and $\ln \Lc=\ln M_{\rm c0}/\msun$, where $M_{\rm c0}$ is the initial
mass of the cluster or $10^6$\,\msun, whatever is smaller). This
relation may remain valid for larger systems built up from many
clusters having masses $\aplt10^6\msun$.  For clusters with $M_{\rm
c0} \apgt 10^7$\,\msun\, we therefore extend the relation as a dashed
line.  The logarithmic factor, however, remains constant, as it refers
to the clusters out of which the bulge formed, not the bulge itself.
The bottom dashed line shows $0.01\Mrunaway$.
%
The five error bars to the left give a summary of the results
presented in Table.\,\ref{Tab:models}; the data are averages of from
left to right: 4k stars (models KML101, KML111 and KML112), 6k (model
6k6X10), 12k (models RxW4 and 12k6X10), 14k (model KML144) and 64k
(model N64R6r36).
The downward pointed arrow gives the upper limit for the mass of a
compact object in the globular cluster M15 (Gebhardt et al.~2000) and
the error bar to the right gives the mass estimate for the compact
object associated with Chandra source \#7 in the irregular Galaxy M82
(Matsumoto \& Tsuru 1999).
The Milky Way is represented by the asterisk using the bulge mass from
Dwek (et al. 1995)\nocite{1995ApJ...445..716D} and the black hole mass
 from Eckart \& Genzel (1997)\nocite{1997MNRAS.284..576E} and Ghez
(2000).\nocite{1998ApJ...509..678G}
Bullets and triangles (upper right) represent the bulge masses and
measured black hole mass of Seyfert galaxies and Quasars, respectively
(both from Wandel 1999;
2001).\nocite{2001astro.ph..8461W}\nocite{1999ApJ...519L..39W} The
dotted lines gives the range in solutions to a least squares fits to
the bullets and triangles (Wandel 2001).
}
\label{fig:bhmass}
\end{figure}

Figure\,\ref{fig:bhmass} shows the relation between the black hole
mass and the bulge mass for several Seyfert galaxies and quasars.  The
expression derived in Sec.\,\ref{sect:ug} and the results of our
\nbody\, calculations (Sec.\,\ref{sect:simulations}) are also
indicated.  The solid and dashed lines (Eq.\,\ref{Eq:mrunaway}) fit
the \nbody\, calculations and enclose the area of the measured black
hole mass--bulge masses.  On the way, the solid curve passes though
two other black-hole mass estimates, for M82 and the globular cluster
M15.  We note that the observed relation between bulge and black hole
masses has a spread of two orders of magnitude.  If this bold
extrapolation really does reflect the formation process of bulges and
central black holes, this spread could be interpreted as a variation
in the efficiency of the runaway merger process.  In that case, only
about one in a hundred star clusters reaches the galactic center,
where its black hole is deposited.

\subsection{Is the globular cluster M15 a special case?}

The possible black hole in the globular cluster M15 may have been
formed by a scenario different from the one described in this paper,
as the cluster's initial relaxation time probably exceeded our upper
limit of 25\,Myr. The current half-mass relaxation time of M15 is
about 2.5\,Gyr (Harris 1996),\nocite{1996yCat.7195....0H} which is far
more than our 100\,Myr limit for forming a massive central object from
a collision runaway.

An alternative is provided by Miller \& Hamilton
(2001),\nocite{2001astro.ph..6188M} who describe the formation of
massive ($\sim 10^3$\,\msun) black holes in star clusters with
relatively long relaxation times.  In their model the black hole grows
very slowly over a Hubble time via occasional collisions with other
stars, in contrast to the model described here, in which the runaway
grows much more rapidly, reaching a characteristic mass of about 0.1\%
of the total birth mass of the cluster within a few megayears.

One possible way around M15's long relaxation time may involve the
cluster's rotation.  Gebhardt (2000; 2001; private communication) has
measured radial velocities of individual stars in the crowded central
field, down to two arcsec of the cluster center.  He finds that, both
in the central part of the cluster ($r < 0.1\rhm$) and outside the
half mass radius, the average rotation velocity is substantial ($\vrot
\apgt 0.5\vdisp$).  Rotation is quickly lost in a cluster, so to
explain a current rotation, M15's initial rotation rate must probably
have been even larger than observed today (see Einsel \& Spurzem
1999).\nocite{1999MNRAS.302...81E} Hachisu (1978; 1982)
\nocite{1979PASJ...31..523H}\nocite{1982PASJ...34..313H} found, using
gaseous cluster models, that an initially rotating cluster tends to
evolve into a 'gravo-gyro catastrophe' which drives the cluster into
core collapse far more rapidly than would occur in a non-rotating
system.  If the gravo-gyro-driven core collapse occurred within
25\,Myr, a collision runaway might have initiated the growth of an
intermediate mass black hole in the core of M15.

\section{Conclusions}

We study the runaway growth of a single star in a dense star cluster
using a combination of complementary approaches.  Our semi-analytic
analysis is supported by detailed \nbody\, calculations in which the
effects of stellar evolution, stellar dynamics, binary evolution and
the perturbing effect of a background Galactic potential are taken
self-consistently into account.

Star clusters with initial half-mass relaxation times $\trlx \aplt
25$\,Myr experience a phase of runaway growth.  In this phase a single
seed star grows to a mass of about 0.1\% of the total mass of the
cluster.  The first collision occurs at the moment the cluster core
collapses. This happens at about 0.2\,\trlx\, but no later than about
5\,Myr (the evolution time scale for a $\apgt 50\msun$ star).  The
star which experiences the first collision becomes the target for
further collisions, initiating runaway growth.  The growth phase is
terminated by (1) the disruption of the cluster in the tidal field of
the Galaxy (at $t\aplt 5\trlx$) or (2) the reversal of core collapse
by mass loss from the evolving stellar population (after about
25\,Myr).

A star cluster can survive for longer than $5\,\trlx$ if, for example,
it did not initially fill its Jacobi surface (``Roche lobe'') in the
Galactic tidal field.  (Examples are NGC\,3603 and R\,136, the dense
star cluster in the 30 Doradus region in the Large Magellanic cloud.)
Such clusters go though a phase of runaway stellar growth, but recover
after stellar mass loss drives the re-expansion of the cluster core.

>From an observational point of view, a tidally limited cluster
experiences three very distinct evolutionary phases: a pre-collapse
phase until 0.2\,\trlx, a phase of deep core collapse (from
0.2\,\trlx\, to about 25\,Myr), followed by an expansion phase
eventually leading to the disruption of the cluster.  During the
expansion phase the cluster half-mass radius expands, causing the mean
relaxation time to increase by a factor of 4 (see Portegies Zwart et
al.\, 2001).  A cluster in this final phase will be observable with a
current relaxation time less than $\sim4\times25\,{\rm Myr} =
100$\,Myr. The clearest indication of its previous phase of core
collapse and runaway growth would be the presence of a central compact
object with a mass $\aplt 0.1$\% of the initial cluster mass.  The
cluster may also be relatively depleted in low-mass compact objects
(stellar mass black holes and neutron stars), as these are consumed
during the runaway growth phase.

Star clusters with an initial relaxation time $\trlx \apgt
25$\,Myr do not experience a phase of runaway growth,
as core collapse is prevented by mass loss from the most massive
stars. These clusters may experience core collapse after $\sim
100$\,Myr, when stellar evolution slows (Takahashi \& Portegies Zwart
1999).\nocite{1998ApJ...503L..49T} This later core collapse, however,
does not lead to a phase of runaway growth. In such old clusters
multiple collisions are still likely to be common and may lead to blue
stragglers with a mass more than twice the turn-off mass.


\acknowledgements 

We are grateful to Karl Gebhardt, Ortwin Gerhard, Pranab Ghosh,
Douglas Heggie, Piet Hut, Jun Makino, and Rainer Spurzem for
discussions and to the anonymous referee for a careful reading of the
manuscript.  We also thank Tokyo University, the Institute for
Advanced Study, the American Museum for Natural History and the
University of Indiana for their hospitality and the use of their
fabulous GRAPE hardware.  This work was supported by NASA ATP grants
NAG5-6964 and NAG5-9264, and by the Nederlands Organization for
Scientific Research (NWO). SPZ was supported as Hubble Fellow
(HF-01112.01-98A) and as fellow of the Royal Netherlands Academy of
Arts and Sciences (KNAW).


\begin{thebibliography}{}

\bibitem[\protect\astroncite{{Antonov}}{1962}]{1962spss.book.....A}
{Antonov}, V.~A. 1962,
\newblock {Solution of the problem of stability of stellar system Emden's
  density law and the spherical distribution of velocities},
\newblock Vestnik Leningradskogo Universiteta, Leningrad: University, 1962

\bibitem[\protect\astroncite{{Binney} \&
  {Tremaine}}{1987}]{1987gady.book.....B}
{Binney}, J., {Tremaine}, S. 1987,
\newblock Galactic dynamics,
\newblock Princeton, NJ, Princeton University Press, 1987, 747 p.

\bibitem[\protect\astroncite{{Chernoff} \&
  {Weinberg}}{1990}]{1990ApJ...351..121C}
{Chernoff}, D.~F., {Weinberg}, M.~D. 1990, \apj, 351, 121

\bibitem[\protect\astroncite{{Cohn}}{1980}]{1980ApJ...242..765C}
{Cohn}, H. 1980, \apj, 242, 765

\bibitem[\protect\astroncite{{Dwek} et~al.}{1995}]{1995ApJ...445..716D}
{Dwek}, E., {Arendt}, R.~G., {Hauser}, M.~G., {Kelsall}, T., {Lisse}, C.~M.,
  {Moseley}, S.~H., {Silverberg}, R.~F., {Sodroski}, T.~J., {Weiland}, J.~L.
  1995, \apj, 445, 716

\bibitem[\protect\astroncite{{Eckart} \& {Genzel}}{1997}]{1997MNRAS.284..576E}
{Eckart}, A., {Genzel}, R. 1997, \mnras, 284, 576

\bibitem[\protect\astroncite{{Einsel} \& {Spurzem}}{1999}]{1999MNRAS.302...81E}
{Einsel}, C., {Spurzem}, R. 1999, \mnras, 302, 81

\bibitem[\protect\astroncite{{Fabbiano} et~al.}{2001}]{2001ApJ...554.1035F}
{Fabbiano}, G., {Zezas}, A., {Murray}, S.~S. 2001, \apj, 554, 1035

\bibitem[\protect\astroncite{{Freitag} \& {Benz}}{2001}]{2001A&A...375..711F}
{Freitag}, M., {Benz}, W. 2001, \aap, 375, 711

\bibitem[\protect\astroncite{{Gebhardt} et~al.}{2000}]{2000AJ....119.1268G}
{Gebhardt}, K., {Pryor}, C., {O'Connell}, R.~D., {Williams}, T.~B., {Hesser},
  J.~E. 2000, \aj, 119, 1268

\bibitem[\protect\astroncite{{Gerhard}}{2001}]{2001ApJ...546L..39G}
{Gerhard}, O. 2001, \apjl, 546, L39

\bibitem[\protect\astroncite{{Gerssen} et~al.}{2001}]{2001AAS...199.5610G}
{Gerssen}, J., {van der Marel}, R.~P., {Dubath}, P., {Gebhardt}, K.,
  {Guhathakurta}, P., {Peterson}, R., {Pryor}, C. 2001, American Astronomical
  Society Meeting

\bibitem[\protect\astroncite{{Ghez} et~al.}{1998}]{1998ApJ...509..678G}
{Ghez}, A.~M., {Klein}, B.~L., {Morris}, M., {Becklin}, E.~E. 1998, \apj, 509,
  678

\bibitem[\protect\astroncite{{Goodman}}{1987}]{1987ApJ...313..576G}
{Goodman}, J. 1987, \apj, 313, 576

\bibitem[\protect\astroncite{{Hachisu}}{1979}]{1979PASJ...31..523H}
{Hachisu}, I. 1979, \pasj, 31, 523

\bibitem[\protect\astroncite{{Hachisu}}{1982}]{1982PASJ...34..313H}
{Hachisu}, I. 1982, \pasj, 34, 313

\bibitem[\protect\astroncite{{Harris}}{1996}]{1996yCat.7195....0H}
{Harris}, W.~E. 1996, VizieR Online Data Catalog

\bibitem[\protect\astroncite{{Heggie} \&
  {Ramamani}}{1995}]{1995MNRAS.272..317H}
{Heggie}, D.~C., {Ramamani}, N. 1995, \mnras, 272, 317

\bibitem[\protect\astroncite{{Hut} \& {Inagaki}}{1985}]{1985ApJ...298..502H}
{Hut}, P., {Inagaki}, S. 1985, \apj, 298, 502

\bibitem[\protect\astroncite{{Joshi} et~al.}{2001}]{2001ApJ...550..691J}
{Joshi}, K.~J., {Nave}, C.~P., {Rasio}, F.~A. 2001, \apj, 550, 691

\bibitem[\protect\astroncite{{Kaaret} et~al.}{2000}]{2000HEAD...32.1511K}
{Kaaret}, P., {Prestwich}, A.~H., {Zezas}, A., {Murray}, S.~S., {Kim}, D.-W.,
  {Kilgard}, R.~E., {Schlegel}, E.~M., {Ward}, M.~J. 2000, AAS/High Energy
  Astrophysics Division, 32, 1511

\bibitem[\protect\astroncite{{Kaaret} et~al.}{2001}]{2001MNRAS.321L..29K}
{Kaaret}, P., {Prestwich}, A.~H., {Zezas}, A., {Murray}, S.~S., {Kim}, D.-W.,
  {Kilgard}, R.~E., {Schlegel}, E.~M., {Ward}, M.~J. 2001, \mnras, 321, L29

\bibitem[\protect\astroncite{{King} et~al.}{2001}]{2001ApJ...552L.109K}
{King}, A.~R., {Davies}, M.~B., {Ward}, M.~J., {Fabbiano}, G., {Elvis}, M.
  2001, \apjl, 552, L109

\bibitem[\protect\astroncite{{King}}{1966}]{1966AJ.....71...64K}
{King}, I.~R. 1966, \aj, 71, 64

\bibitem[\protect\astroncite{{Kotoku} et~al.}{2000}]{2000PASJ...52.1081K}
{Kotoku}, J., {Mizuno}, T., {Kubota}, A., {Makishima}, K. 2000, \pasj, 52, 1081

\bibitem[\protect\astroncite{{Lee}}{1987}]{1987ApJ...319..801L}
{Lee}, H.~M. 1987, \apj, 319, 801

\bibitem[\protect\astroncite{{Makino}}{1996}]{1996ApJ...471..796M}
{Makino}, J. 1996, \apj, 471, 796

\bibitem[\protect\astroncite{{Makino}}{2001}]{2001dscm.conf...87M}
{Makino}, J. 2001,
\newblock in ASP Conf. Ser. 228: Dynamics of Star Clusters and the Milky Way,
  ~87

\bibitem[\protect\astroncite{{Makino} \& {Hut}}{1990}]{1990ApJ...365..208M}
{Makino}, J., {Hut}, P. 1990, \apj, 365, 208

\bibitem[\protect\astroncite{{Makino} et~al.}{1997}]{1997ApJ...480..432M}
{Makino}, J., {Taiji}, M., {Ebisuzaki}, T., {Sugimoto}, D. 1997, \apj, 480, 432

\bibitem[\protect\astroncite{{Matsumoto} et~al.}{2000}]{2000HEAD...32.0108M}
{Matsumoto}, H., {Canizares}, C.~R., {Tsuru}, T.~G., {Koyama}, K., {Awaki}, H.,
  {Matsushita}, S., {Prestwitch}, A., {Zezas}, A.~L., {Kawai}, N., {Ward}, M.,
  {Kawabe}, R. 2000, AAS/High Energy Astrophysics Division, 32, 0108

\bibitem[\protect\astroncite{{Matsumoto} et~al.}{2001}]{2001ApJ...547L..25M}
{Matsumoto}, H., {Tsuru}, T.~G., {Koyama}, K., {Awaki}, H., {Canizares}, C.~R.,
  {Kawai}, N., {Matsushita}, S., {Kawabe}, R. 2001, \apjl, 547, L25

\bibitem[\protect\astroncite{{Matsushita} et~al.}{2000}]{2000ApJ...545L.107M}
{Matsushita}, S., {Kawabe}, R., {Matsumoto}, H., {Tsuru}, T.~G., {Kohno}, K.,
  {Morita}, K., {Okumura}, S.~K., {Vila-Vilar{\' o}}, B. 2000, \apjl, 545, L107

\bibitem[\protect\astroncite{{McMillan}}{1986}]{1986ApJ...306..552M}
{McMillan}, S. L.~W. 1986, \apj, 306, 552

\bibitem[\protect\astroncite{{Mengel} et~al.}{2001}]{2001ApJ...550..280M}
{Mengel}, S., {Lehnert}, M.~D., {Thatte}, N., {Tacconi-Garman}, L.~E.,
  {Genzel}, R. 2001, \apj, 550, 280

\bibitem[\protect\astroncite{{Meynet} et~al.}{1994}]{1994A&AS..103...97M}
{Meynet}, G., {Maeder}, A., {Schaller}, G., {Schaerer}, D., {Charbonnel}, C.
  1994, \aaps, 103, 97

\bibitem[\protect\astroncite{{Mezger} et~al.}{1996}]{1996A&ARv...7..289M}
{Mezger}, P.~G., {Duschl}, W.~J., {Zylka}, R. 1996, \aapr, 7, 289

\bibitem[\protect\astroncite{{Miller} \&
  {Hamilton}}{2001}]{2001astro.ph..6188M}
{Miller}, C.~M., {Hamilton}, D.~P. 2001,
\newblock in {astro-ph/0106188}

\bibitem[\protect\astroncite{{Mizuno} et~al.}{1999a}]{1999AN....320..356M}
{Mizuno}, T., {Ohnishi}, T., {Kubota}, A., {Makishima}, K., {Tashiro}, M.
  1999a, Astronomische Nachrichten, 320, 356

\bibitem[\protect\astroncite{{Mizuno} et~al.}{1999b}]{1999PASJ...51..663M}
{Mizuno}, T., {Ohnishi}, T., {Kubota}, A., {Makishima}, K., {Tashiro}, M.
  1999b, \pasj, 51, 663

\bibitem[\protect\astroncite{{Portegies Zwart} et~al.}{1999}]{pzmmh99}
{Portegies Zwart}, S.~F., {Makino}, J., {McMillan}, S. L.~W., {Hut}, P. 1999,
  \aap, 348, 117

\bibitem[\protect\astroncite{{Portegies Zwart}
  et~al.}{2001a}]{2001ApJ...546L.101P}
{Portegies Zwart}, S.~F., {Makino}, J., {McMillan}, S. L.~W., {Hut}, P. 2001a,
  \apjl, 546, L101

\bibitem[\protect\astroncite{{Portegies Zwart} \&
  {McMillan}}{2000}]{2000ApJ...528L..17P}
{Portegies Zwart}, S.~F., {McMillan}, S. L.~W. 2000, \apjl, 528, L17

\bibitem[\protect\astroncite{{Portegies Zwart}
  et~al.}{2001b}]{2001MNRAS.321..199P}
{Portegies Zwart}, S.~F., {McMillan}, S. L.~W., {Hut}, P., {Makino}, J. 2001b,
  \mnras, 321, 199

\bibitem[\protect\astroncite{{Quinlan} \&
  {Shapiro}}{1990}]{1990ApJ...356..483Q}
{Quinlan}, G.~D., {Shapiro}, S.~L. 1990, \apj, 356, 483

\bibitem[\protect\astroncite{{Salpeter}}{1955}]{1955ApJ...121..161S}
{Salpeter}, E.~E. 1955, \apj, 121, 161

\bibitem[\protect\astroncite{{Sanders}}{1970}]{1970ApJ...162..791S}
{Sanders}, R.~H. 1970, \apj, 162, 791

\bibitem[\protect\astroncite{{Sanders} \&
  {Lowinger}}{1972}]{1972AJ.....77..292S}
{Sanders}, R.~H., {Lowinger}, T. 1972, \aj, 77, 292

\bibitem[\protect\astroncite{Scalo}{1986}]{scalo86}
Scalo, J.~M. 1986, Fund. of Cosm. Phys., 11, 1

\bibitem[\protect\astroncite{{Spitzer}}{1987}]{1987degc.book.....S}
{Spitzer}, L. 1987,
\newblock Dynamical evolution of globular clusters,
\newblock Princeton, NJ, Princeton University Press, 1987, p. 191

\bibitem[\protect\astroncite{{Spitzer}}{1969}]{1969ApJ...158L.139S}
{Spitzer}, L.~J. 1969, \apjl, 158, L139

\bibitem[\protect\astroncite{{Spitzer} \& {Hart}}{1971}]{1971ApJ...164..399S}
{Spitzer}, L.~J., {Hart}, M.~H. 1971, \apj, 164, 399

\bibitem[\protect\astroncite{{Takahashi} \& {Portegies
  Zwart}}{1998}]{1998ApJ...503L..49T}
{Takahashi}, K., {Portegies Zwart}, S.~F. 1998, \apjl, 503, L49

\bibitem[\protect\astroncite{{Vishniac}}{1978}]{1978ApJ...223..986V}
{Vishniac}, E.~T. 1978, \apj, 223, 986

\bibitem[\protect\astroncite{{Wandel}}{1999}]{1999ApJ...519L..39W}
{Wandel}, A. 1999, \apjl, 519, L39

\bibitem[\protect\astroncite{{Wandel}}{2001}]{2001astro.ph..8461W}
{Wandel}, A. 2001,
\newblock in {astro-ph/0108461},  8461

\bibitem[\protect\astroncite{{Zezas} \& {Fabbiano}}{2002}]{2002astro.ph..3176Z}
{Zezas}, A., {Fabbiano}, G. 2002,
\newblock in {astro-ph/0202176}

\bibitem[\protect\astroncite{{Zezas} et~al.}{2002}]{2002astro.ph..3174Z}
{Zezas}, A., {Fabbiano}, G., {Rots}, A.~H., {Murray}, S.~S. 2002,
\newblock in {astro-ph/0203174}

\end{thebibliography}
\end{document}